\documentstyle[prl,aps,twocolumn]{revtex}

\input epsf

\begin{document}

\wideabs{
\title
{
Bosons in cigar-shape traps: Thomas-Fermi regime, Tonks-Girardeau regime, and between 
}
\author
{
V. Dunjko$^{1}$, V. Lorent$^{2}$, and M. Olshanii$^{1,2}$
\cite{e-mail}
}
\address
{
$^{1}$Department of Physics \& Astronomy, 
University of Southern California, Los Angeles, California 90089-0484, USA
\\
$^{2}$Laboratoire de Physique des Lasers - Institut Galil\'{e}e,
Universit\'{e} Paris-Nord, F-93430 Villetaneuse, France
}

\date{\today}
\maketitle
\begin{abstract}
We present a quantitative
analysis of the experimental accessibility of the
Tonks-Girardeau gas in the current day experiments with cigar-trapped
alkalis.
For this purpose
we derive,
using a Bethe anzats generated local equation of state,
a set of hydrostatic equations describing
one-dimensional $\delta$-interacting Bose gases
trapped in a harmonic potential.
The resulting solutions
cover the  {\it entire range} of atomic densities.
\end{abstract}
}

Physics of trapped atomic gases attracted an enormous 
attention of both experimentalists and theorists in the last 
decade \cite{Eric-Wolfgang-Hulett,Stringari}. All atomic systems 
experimentally accessed so far 
can be well understood within a so called 
mean-field picture where the atom-atom correlations are 
weak and a particular atom can be effectively viewed as 
moving in a mean-field created by others and unaffected     
by the presence of the probe atom. Small deviations 
from the mean-field predictions can be 
calculated using the Bogoliubov approximation 
\cite{Bogoliubov}. An interesting question arrises as what 
happens beyond the scope of the mean-field behavior. 
For the three-dimensional systems the far-beyond-mean-field 
regime has neither been reached experimentally (the JILA
experimental group has made a significant progress towards this goal
\cite{JILA_resoneum}), nor understood theoretically. 

For quasi-one-dimensional
systems (cigar-shape traps) the situation is different. Recall that 
a one-dimensional system of zero-range interacting bosons 
in a flat-bottom box
is exactly integrable \cite{Lieb} 
via Bethe anzats, for {\it all} values of the coupling 
strength. The exactly known equation of state in a box provides
an input for a unified beyond-mean-field treatment 
of trapped quasi-one-dimensional atomic gases, and that 
is what our paper is devoted to. The main goal of the paper 
is to provide quantitative criteria for an experimental realization of the 
Tonks-Girardeau gas \cite{Tonks-Girardeau1} 
in atomic experiments, with 
the main emphasis on deviations of the spatial distribution
from the mean-field Thomas-Fermi profile \cite{Baym}. 
We paid special attention to the intermediate between
weak and strong interactions regime, as the most realistic from the experimental point of view. 
The extreme case of infinitely strongly interacting atoms  
in a harmonic potential has been already 
investigated in Ref.\cite{Kolomeisky-Girardeau2}.
Also the beyond-mean-field
effects in the self-correlation function of the trapped one-dimensional gas were considered in
\cite{Gora}.  

Experimental progress in creating one-dimensional atomic gases is also quite fast. 
One-dimensional guiding of thermal cold atoms through elongated magnetic configurations \cite{Nynke}
has already been successfully demonstarted. A recent
experiment \cite{Ertmer} shows evidence of trapping a Bose-Einstein condensate by a shallow magnetic gradient in one
direction combined with a blue detuned hollow laser beam ensuring tight confinement in the radial directions.
To ensure a truly one-dimensional behavior the atoms must be cold enough to predominantly occupy 
the ground transverse vibrational state. 
In this respect the most promising candidate is atoms in an  array of 1D traps formed in the  intersection of two
far-detuned Gaussian standing waves \cite{Weiss_private}, obtained in turn by releasing  a full, three-standing-waves 
optical lattice, where up to $\sim$20\% ground vibrational mode occupation is already experimentally accessible 
\cite{Weiss_vibrational}. A 2D version 
of such an array (one standing wave) has been already reported in literature \cite{Weiss_2D}.      

In highly elongated cigar-shaped traps ($\omega_{\bot} \gg \omega_{z}$), 
the transverse atomic motion is 
governed by the Hamiltonian
\begin{eqnarray}
\widehat{H}_{\bot}
=
\frac{\widehat{p}^{2}_{x}+\widehat{p}^{2}_{y}}{2 m}
+
\frac{m \omega^{2}_{\bot}\left(x^{2}+y^{2}\right)}{2}\,\,\,, 
\label{transH}
\end{eqnarray}
where $m$ is the mass of the atoms, and $\omega_{\bot}$ and $\omega_{z}$ are 
the transverse and longitudinal frequencies of the trap respectively.
If 
neither temperature nor interaction energy-per-particle $\epsilon$
(defined below) exceed the transverse level spacing $\hbar\omega_{\bot}$,
atoms 
occupy the ground mode of this Hamiltonian, 
and the system becomes effectively one-dimensional \cite{Olshanii1}. In
particular, the interparticle interaction in the longitudinal direction 
can be well-approximated by an effective two-atom interaction potential
\cite{Olshanii1}
\begin{equation}
U_{\rm 1D}(z)=g_{\rm 1D}\,\delta(z) \,\,\, .
\label{1D_potential}
\end{equation}
where
$
g_{\rm 1D}=-\hbar^{2}/\tilde{\mu}\,a_{\rm 1D},
$
is an
effective one-dimensional coupling constant, and 
$
a_{\rm 1D}
=
(-a^{2}_{\bot}/2 a)\,[1-{\mathcal C}(a/a_{\bot})]
$
is the one-dimensional scattering length  
defined analogously to the three-dimensional case as
$a_{\rm 1D} = -\partial\Delta/\partial k_{z} |_{k_{z} \to 0^{+}}$, 
$\Delta(k_{z})$ being the scattering phase of the even scattered
wave. Here and below $\tilde{\mu} = m/2$ is the reduced mass,
$
a_{\bot}=\sqrt{\hbar / \tilde{\mu} \,\omega_{\bot}}
$ is the size of the ground state of the transverse
Hamiltonian in Eq. (\ref{transH}), $a$ is the three-dimensional
scattering length, and 
$
{\mathcal C} =\, \lim_{s
\to \infty}\left(\int^{s}_{0} \, ds'/\sqrt{s'} \,\,
- \,\, \sum_{s'=1}^{s} 1/\sqrt{s'} \right) = 1.4603
\dots \,
$.

Up to terms containing $\mathcal C$,
the effective one-dimensional interaction is a simple projection
of the three-dimensional zero-range interaction on the transverse ground mode. 
Transverse renormalization 
effects captured in the $\mathcal C$-dependent terms become 
important only for  
strong confinement, $\left| a \right| \gg a_{\bot}$.

In addition to transverse trapping and interparticle interactions, the
atoms are also subject to weak residual longitudinal trapping, which
is usually represented by a one-dimensional harmonic potential of 
a frequency $\omega_{z}$. The effective one-dimensional Hamiltonian for
$N$ trapped atoms is  thus
\begin{equation}
\widehat{H}_{\rm 1D}
=
\widehat{H}_{\rm 1D}^{0} + \sum_{j=i}^{N}\,\frac{m\omega_{z}^{2} z_{i}^{2}}{2}
\,\,\,,
\end{equation}
where
\begin{equation}
\widehat{H}_{\rm 1D}^{0}=
-\frac{\hbar^{2}}{2m}\sum_{j=1}^{N}\,\frac{\partial^{2}}{\partial z_{j}^{2}}
+ 
g_{\rm 1D}\sum_{i=1}^{N} \, \sum_{j=i+1}^{N} \, \delta \left(z_{i}-z_{j}\right)
\end{equation}
is the well-known Hamiltonian for a one-dimensional $\delta$-interacting Bose
gas in a flat-bottom box. 
This Hamiltonian can be diagonalized via Bethe anzats \cite{Lieb},
and thus  
the equation of state of such a gas is known
{\it exactly} for all densities and temperatures \cite{Yang-Yang}. 
At zero temperature the energy per particle
$\epsilon(n)$ 
is given through 
\begin{equation}
\epsilon(n)= \frac{\hbar^{2}}{2m}n^{2} e(\gamma(n)) 
\label{epsilon} \,\,\,,
\end{equation}
where the dimensionless parameter 
$
\gamma = 2/n|a_{\rm 1D}|
$
is inversely proportional to the one-dimensional 
gas parameter $n \left|a_{\rm 1D}\right|$, $n$ is
the one-dimensional number density of particles,
and the function $e(\gamma)$ is
given by
\begin{equation}
e(\gamma) = \frac{\gamma^{3}}{\lambda^{3}(\gamma)}
\int_{-1}^{1}\,g\left(x|\gamma\right)x^{2}\,dx ,
\end{equation}
Functions  $g\left(x|\gamma\right)$ and $\lambda(\gamma)$ are solutions of
the Lieb-Liniger system of equations \cite{Lieb}:
\begin{eqnarray}
g\left(x|\gamma\right)-\frac{1}{2\pi}\int_{-1}^{1}\,
\frac{2\lambda(\gamma)}{\lambda^{2}(\gamma)+(y-x)^{2}}\,g\left(y|\gamma\right)
\, dy=\frac{1}{2\pi} 
\label{system1}
\\
\lambda(\gamma)=\gamma \int_{-1}^{1}\, g\left(x|\gamma\right) \, dx \,\,\,.
\label{system2}
\end{eqnarray}

Limits of small and large gas parameter $n\left|a_{\rm 1D}\right|$ 
can be expressed in closed form (see \cite{Lieb}):
\begin{eqnarray} 
\mbox{\rm as} \;n\left|a_{\rm 1D}\right| \to 0,  \quad  \epsilon(n) \to
\frac{\pi^{2}\hbar^2}{6m} \,n^{2} \quad; 
\label{Tnk} 
\\
\mbox{\rm as} \;n\left|a_{\rm 1D}\right| \to \infty,  \quad \epsilon(n) \to \frac{1}{2}\,
g_{\rm 1D}\, n \quad.  
\label{ThFe}
\end{eqnarray}
The low-density limit (\ref{Tnk}) corresponds to the case of 
infinitely strong interactions; the corresponding atomic system 
is usually referred as a gas of impenetrable bosons or Tonks-Girardeau gas
\cite{Tonks-Girardeau1}. Notice that the 
expression for the energy-per-particle
formally coincides with the one for free fermions: 
this is a manifestation of the Fermi-Bose duality in 
one-dimensional systems \cite{duality}. 
The opposite 
limit of high densities (\ref{ThFe}) represents (if multiplied by the 
number of particles and integrated over volume)  the Thomas-Fermi energy 
functional first introduced by Bogoliubov \cite{Bogoliubov}.  
It is usually viewed as the thermodynamic limit of the 
Gross-Pitaevskii energy functional \cite{Gross-Pitaevskii},
derived in turn on the basis of the 
mean-field approximation. It should be mentioned that 
in the three-dimensional case the mean-field approximation is valid 
for low densities contrary to the one-dimensional case where it is valid 
at high densities instead.        

We now introduce the classical hydrodynamical approximation: we
suppose that the trapped gas at each position $z$ is in local
thermal equilibrium, with local
energy per particle given through Eqn. (\ref{epsilon}).
At zero temperature the hydrodynamic equations of motion read
%
\begin{eqnarray}
&&
\frac{\partial}{\partial t}n + \frac{\partial}{\partial z}(n v)
=
0
\nonumber
\\
&&
\frac{\partial}{\partial t} v + v \frac{\partial}{\partial z} v
=
-(1/m) \frac{\partial}{\partial z} (\phi(n) + V(z))
\,\,\, ,
\label{hydrodynamics}
\end{eqnarray}
where $v$ is the atomic velocity,
$V(z)=m \omega_{z}^{2} z^{2}/2$ is the potential energy per
particle, and
\begin{eqnarray}
\phi(n) \stackrel{T=0}{=} (1 + n\, \frac{\partial}{\partial n}) \, \epsilon(n)
\label{phi}
\end{eqnarray}
is the Gibbs free energy per particle obtained using the 
Eqns.~\ref{epsilon}-\ref{system2} and \ref{hydrodynamics}-\ref{phi} 
and tabulated in   
Ref.~\cite{table}. The validity
of the hydrodynamical approximation requires the
typical energy per particle $\epsilon \sim mv^2/2 + \phi + V$
be much higher than the longitudinal quantum level spacing $\hbar\omega_{z}$.

In what follows we will be interested in a {\it steady-state}
solution of the system (\ref{hydrodynamics}). The first integral
of the stationary version of this system obviously reads
\begin{eqnarray}
\begin{array}{ll}
\phi(n) + V(z) = \mu  & \quad,\quad \mbox{for } |z| \le R
\\
n = 0                 & \quad,\quad \mbox{for } |z| > R
\end{array}
\label{intEul}
\end{eqnarray}
where $\mu$ can be proven to be the chemical potential of the system,
fixed by a normalization condition
\begin{equation}
\int_{-R}^{+R} \, n(z) \, dz = N \,\,\,.
\label{Ntot}
\end{equation}
Here the atomic cloud radius $R$ is given by
$V(R) = \mu$.
\begin{figure}
\begin{center}
\leavevmode \epsfxsize=0.45\textwidth 
\epsfbox{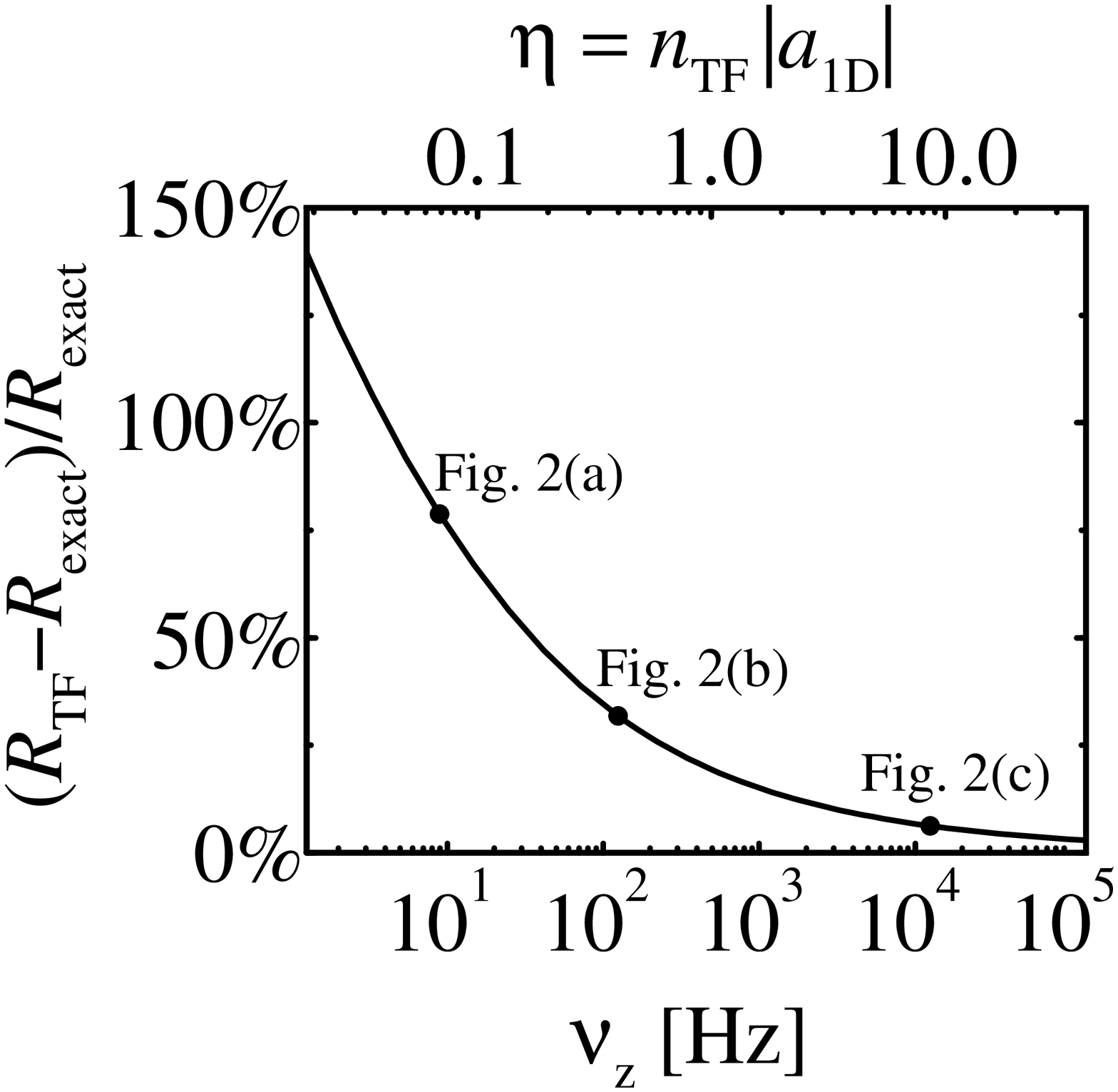} 
\end{center}
\caption{Discrepancy between the cloud size computed from the exact
equation of state and from the Thomas-Fermi prediction, as a function
of the (single) governing parameter $\eta$ (\ref{eta}) shown 
in the upper $x$-axis. 
As an example of a ``physical'' variation of the parameter $\eta$
we consider $N=200$ 
confined in a cigar-shape trap
of transverse frequency $\nu_{\bot} = 180 \,\mbox{kHz}$
$^{133}$Cs atoms, whose scattering length 
is magnetically tuned [20] to $a=+120 \, a_{{\rm Bohr}}$. 
Variation of the 
parameter $\eta$ is supposed to be realized through the 
variation of the longitudinal frequency $\nu_{z}$ (lower $x$-axis).  
For the three representative points marked in this Figure,
we computed the density profiles from the exact, Thomas-Fermi and
Tonks-Girardeau equations of state. Results are shown in Fig. \ref{nprofile}.  
}
\label{RvsRTF}
\end{figure}
In the limit of low density (\ref{Tnk}) the density profile is
a square root of parabola \cite{Kolomeisky-Girardeau2}
\begin{equation}
n(z)=n^{0}_{\rm Tonks}\,\left(1-\frac{z^{2}}{R_{\rm Tonks}^{2}}\right)^{1/2}
\label{Tonks_profile}
\end{equation}
for $z \in [-R_{\rm Tonks},R_{\rm Tonks}]$, with $n=0$ elsewhere, where
\begin{eqnarray}
&&
n^{0}_{\rm Tonks}
=
\left[(2/\pi^2)\, N\, (m\omega_{z}/\hbar)\right]^{1/2} \quad ;
\\
&&R_{\rm Tonks} = \left[2\, N\, (\hbar/\omega_{z}m) \right]^{1/2} \quad .
\end{eqnarray}
The opposite limit (\ref{ThFe}) reproduces the familiar Thomas-Fermi
parabola \cite{Baym}. In this limit, it is easy to show that
\begin{equation}
n(z)=n^{0}_{\rm TF}\,\left(1-\frac{z^{2}}{R_{\rm TF}^{2}}\right)
\label{TF_profile}
\end{equation}
for $z \in [-R_{\rm TF},R_{\rm TF}]$, with $n=0$ elsewhere, where
\begin{eqnarray}
&&
n^{0}_{\rm TF}
=
\left[(9/64)\, N^{2}\, \left(m\omega_{z}/\hbar\right)^{2}
\left|a_{\rm 1D}\right|\right]^{\frac{1}{3}} \quad ;
\\
&&
R_{\rm TF}=\left[3\,N\, \left(\hbar/m \omega_{z}\right)^{2}
/\left|a_{\rm 1D}\right|\right]^{\frac{1}{3}} \quad .
\label{RTF}
\end{eqnarray}
In general, the rescaled density $s=n/n^{0}_{\rm TF}$
satisfies the rescaled version of the system
(\ref{intEul}) and (\ref{Ntot}),
\begin{eqnarray}
\frac{1}{2\,\eta}
\,f \left( \frac{2}{\eta \, s(x)}
    \right)
+
\frac{1}{2}x^{2}
=
\frac{b^{2}}{2}
&&
\label{dim_less_1}
\\
\int_{-b}^{b}\,s(x)\,dx = \frac{4}{3}\,\,\,,
\label{dim_less_2}
&&
\end{eqnarray}
with $x=z/R_{\rm TF}$ is the dimensionless coordinate, and
$
f(\gamma)=\left[3e(\gamma)-\gamma
de/d\gamma\right]/\gamma^{2}
$
is the dimensionless Gibbs energy
(see \cite{table}). We see that this system has only one
governing parameter,
\begin{eqnarray}
\eta
&\equiv&
n^{0}_{\rm TF}\left|a_{\rm 1D}\right|
\label{eta}
\\
&=&
\frac{(9/2)^{1/3}}{8} \,
\left(
\frac
 { a_{\bot}^4 m N \omega_{z} (1-{\cal C} a/a_{\bot})^2 }
 {a^2 \hbar}
\right)^{2/3}
\,\,\,,
\nonumber
\end{eqnarray}
whose physical meaning is the Thomas-Fermi estimate
of the one-dimensional gas parameter in the center of the trap.
It is easy to show that the limit of
small $\eta$ corresponds to the Tonks-Girardeau limit,
and the limit of large $\eta$, to the
Thomas-Fermi limit:
\begin{eqnarray}
\begin{array}{lcl}
\eta
\ll 1
&
\longrightarrow
&
\mbox{\rm Tonks-Girardeau profile (\ref{Tonks_profile})}
\\
\eta
\gg 1
&
\longrightarrow
&
\mbox{\rm Thomas-Fermi profile (\ref{TF_profile})} \quad .
\end{array}
\end{eqnarray}
\begin{figure}
   \leavevmode
   \centering
   \parbox{4cm}
   {  
   \begin{center}
   \epsfxsize=0.22\textwidth
                                                         \epsfbox{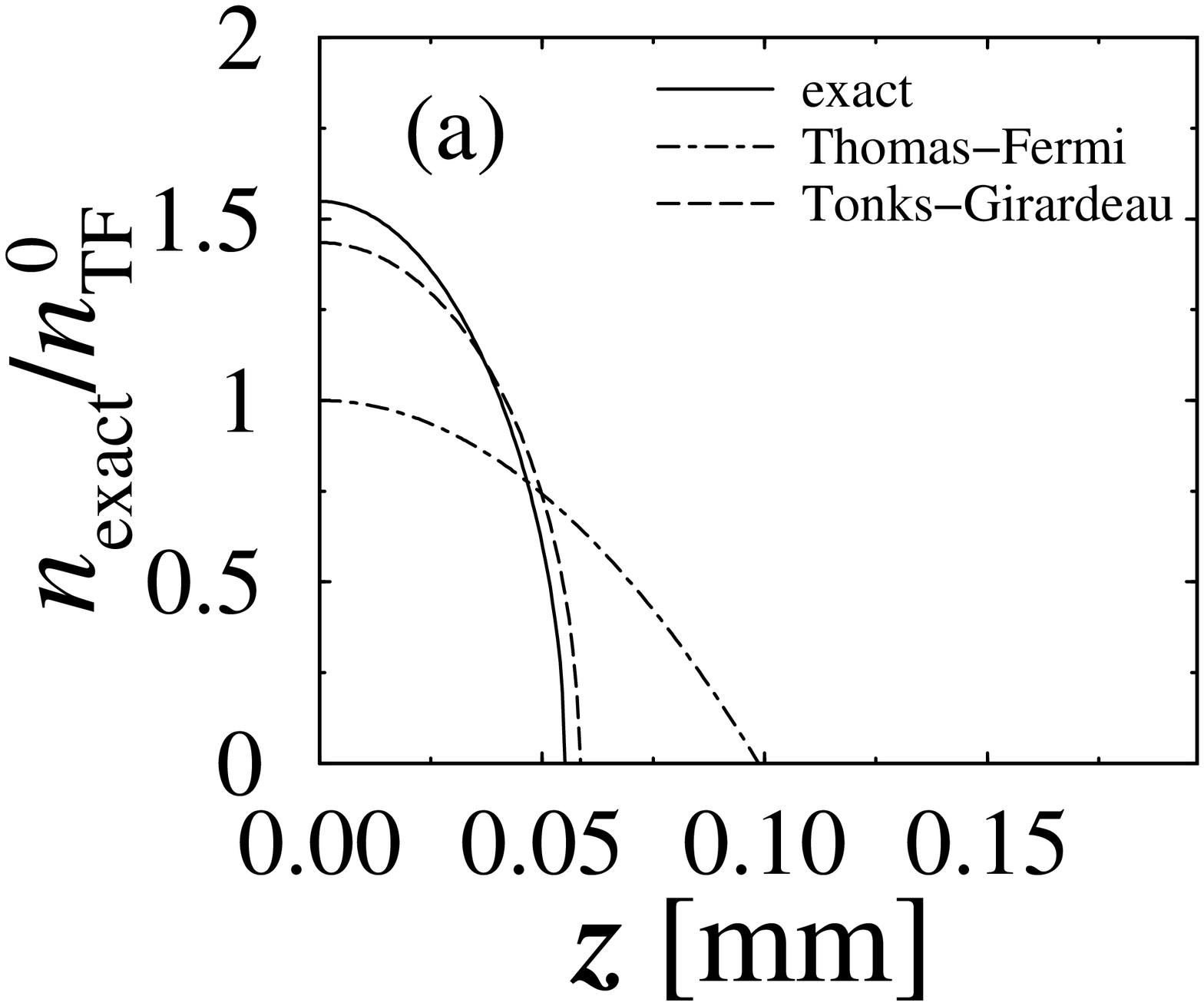}
   \end{center}
   }  
   \parbox{4cm} 
   {  
   \begin{center} 
   \epsfxsize=0.22\textwidth
                                                         \epsfbox{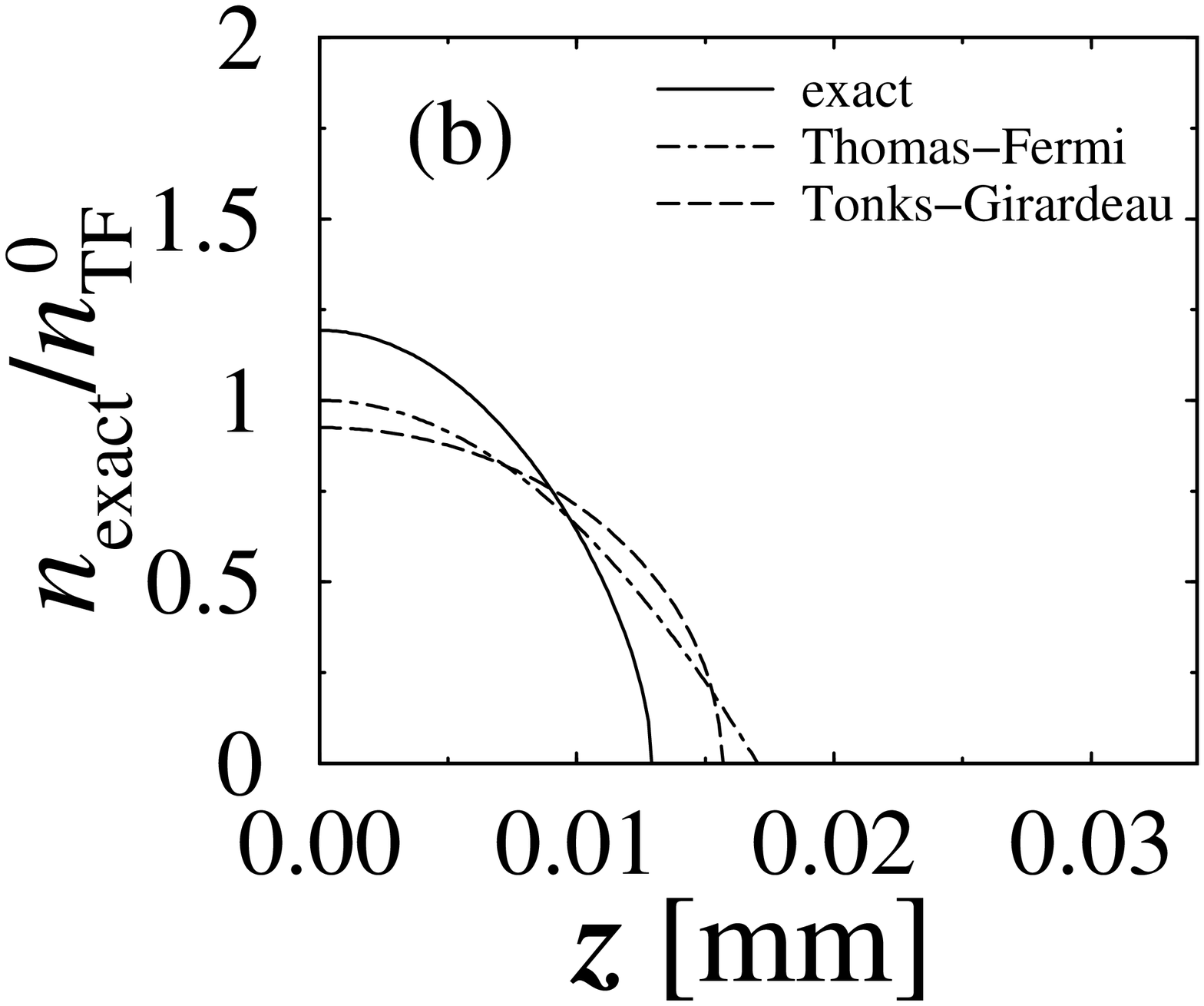}
   \end{center}
   }  
   \parbox{4cm}  
   {  
   \begin{center}  
   \epsfxsize=0.22\textwidth
                                                         \epsfbox{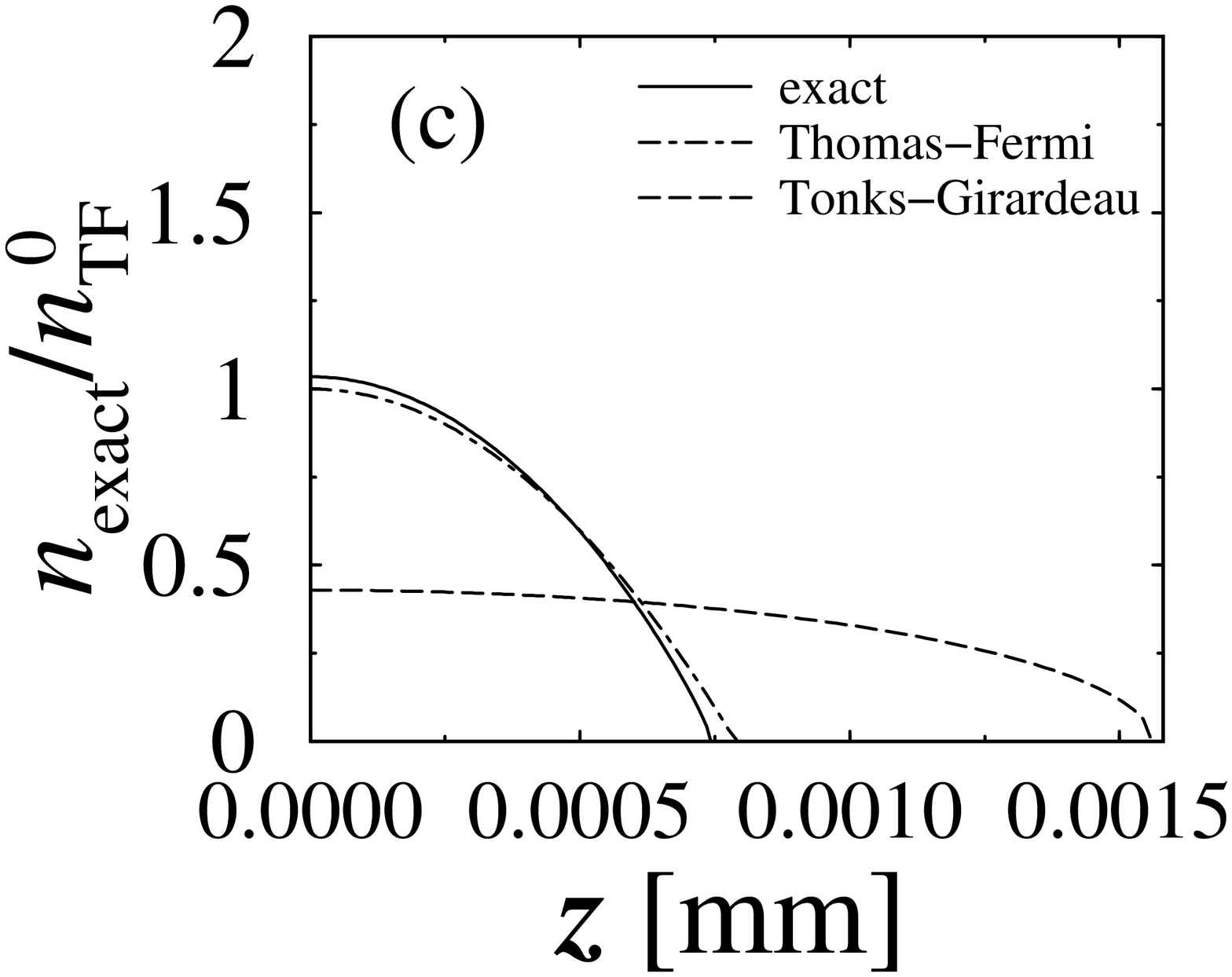}
   \end{center} 
   }  
\caption
{ Comparison of density profiles produced by Thomas-Fermi, exact, and
Tonks-Girardeau local equations of state for the points indicated in
Fig. \ref{RvsRTF}. (a) $\eta = 0.07$,
$\nu_{z}$ = 10 Hz; Tonks-Girardeau regime. (b) $\eta = .4$,
$\nu_{z}$ = 130 Hz; intermediate regime. (c)
$\eta = 9$, $\nu_{z}$ = 13 kHz;
Thomas-Fermi regime. The number of atoms, three-dimensional scattering
length, and transverse frequency are the same as in Fig. \ref{RvsRTF}.
}
\label{nprofile}
\end{figure}
The exactly known equation of state
Eqs. (\ref{epsilon}) allows one to precisely map
the transition between these two limiting behaviors. A convenient way
to detect, both experimentally and numerically, the degree to which
the system is in one or the other limiting regime is to compare the
size of the cloud of trapped atoms, $R$ computed from the Eqs. (\ref{intEul},\ref{Ntot}), 
to the
Thomas-Fermi prediction $R_{\rm TF}$ in Eq. (\ref{RTF}). This is shown
in Fig. \ref{RvsRTF}. Density profiles for three representative values of 
the governing parameter $\eta$ 
are plotted in
Fig. \ref{nprofile}. 
As the Figs.\ref{RvsRTF}-\ref{nprofile} make clear,
the transition between the two limiting regimes happens over the range
of the governing parameter $\eta$ 
from about 0.1 to 10, reaching 75\% of the cloud size discrepancy for the former. 

Let us summarize the requirements for a successful observation of the 
zero-temperature Tonks-Girardeau density profile (\ref{Tonks_profile}):
\begin{itemize}
\item[(a)] The coupling constant $g_{\rm 1D}$ (see (\ref{1D_potential}))
must be positive. This 
condition is equivalent to $0 <a < a_{\bot}/{\cal C}$;
\item[(b)] 
To ensure the  
one-dimensional behavior of the system the energy per particle 
must not exceed the transverse 
level spacing: in the case of the Tonks-Girardeau gas this condition  
leads to  
$\hbar\omega_{z} N \ll \hbar\omega_{\bot}$;
\item[(c)]
The number of particles must be well below the Tonks-Girardeau vs.\ Thomas-Fermi 
boundary which requires $\eta \ll 1$, where $\eta$ is given
by (\ref{eta});
\item[(d)] It turns out that not for every set of 
trap and atom parameters the energy per particle at the Tonks-Girardeau vs.\ Thomas-Fermi
boarder ($\eta \sim 1$) is consistent with the hydrodynamical approximation, {\it i.e.\ }exceeds
the longitudinal
level spacing. To avoid that 
one must require 
$\hbar^2/m \left|a_{\rm 1D}\right|^2 \gg \hbar\omega_{z}$  or 
$a  \gg (a_{\bot}^2/a_{z})(1-{\cal C} a/a_{\bot})$,
where $a_{z} \sim (\hbar/m\omega_{\bot})^{1/2}$ 
is the size of the longitudinal ground state. If the above condition 
is violated, then, as the number of particles decreases,
the system will pass from the Thomas-Fermi regime to the ideal gas directly,
skipping the Tonks-Girardeau regime;
\item[(f)] 
One may require also, that when already in the Tonks-Girardeau regime ($\eta \ll 1$), 
the number of atoms must nevertheless be high enough to ensure the 
hydrodynamical, non-ideal gas behavior. It turns out that this requirement leads to 
a trivial condition 
of having more than one atom in the system: $N \gg 1$. 
\end{itemize}
%

%
{\bf Acknowledgments}. Authors are grateful to D.~S.~ Weiss and 
V.~V.~Kresin
for enlightening discussions on the subject. This work was supported by
the NSF grant {\it PHY-0070333}.
\end{document}